\documentclass[twocolumn]{article}
\usepackage{amssymb,amsmath}
\usepackage{relsize}
\usepackage[utf8]{inputenc}
\usepackage{graphicx}
\usepackage{verbatim}

\title{From Particle Tracks to Velocity and Acceleration Fields Using B-Splines and Penalties}

\author{Sebastian Gesemann\\ Experimental Methods, DLR, Göttingen, Germany}

\begin{document}

\maketitle

\begin{abstract}
In this work a method for reconstructing velocity and acceleration fields is described which uses scattered particle tracking data from flow experiments as input. The goal is to reconstruct these fields faithfully with a limited amount of compute time and exploit known flow properties such as a divergence-free velocity field for incompressible flows and a rotation-free acceleration in case it is known to be dominated by the pressure gradient in order to improve the spatial resolution of the reconstruction.
\end{abstract}

\section{Introduction}

Determining 3D velocity fields from flow experiments is difficult especially if a high
spatial resolution is desired. For optical particle-based measurement techniques, a high spatial resolution implies a high density of tracer particles that need follow the flow and have to be observed using multiple cameras.

Instead of identifying and matching particles separately between different views, which is nontrivial at high particle densities, a tomographic approach has been successfully applied in the past to the particle distribution reconstruction problem. This method is called \emph{TomoPIV} for Tomographic Particle Image Velocimetry. In TomoPIV the measurement volume is discretized and reconstructed as the solution to a large but sparse and constrained linear equation system resulting in a discrete volume of light intensities. One way of deriving flow fields from these volumes is to apply a cross correlation between two subvolumes of reconstructions from neighbouring points of time to detect the average flow velocity within such a subvolume. This is a robust method to compute flow velocities but results in a spatially lowpass filtered representation of the velocity depending on the size of the subvolumes used for cross-correlation usually called window size.

With recent advances in particle tracking techniques the density of particles that can be reconstructed directly without discretizing the volume is approaching densities that are typically used in TomoPIV measurements. The advantage of methods for reconstructing particle locations directly instead of a discretized volume has several benefits over a tomographic reconstruction: It typically requires a only small fraction of CPU and RAM resources compared to what is needed for TomoPIV to solve the large constrained equation system. Also, such a direct particle reconstruction method avoids an additional layer of spatial discretization which can be expected to improve the accuracy of particle location measurements. Given a sequence of time-resolved measurement images of a flow with tracer particles, particle tracks can be reconstructed with these new techniques.

In this publication, we will describe a method for reconstructing velocity and acceleration fields from scattered and noisy particle tracks which we developed with the goal of preserving much of the information present in the particle tracks and avoiding any unwanted spatial lowpass filtering effect such as the one that is inherent in correlation-based methods. In addition, it's possible to exploit prior knowledge to improve the spatial resolution of the reconstructed field such as as freedom of divergence of the velocity field for cases with incompressible flow.

\section{Overview}

Our approach to compute velocity and acceleration fields based on noisy particle tracks can be split into two parts:

\begin{itemize}
\item \emph{trackfit} takes the noisy particle location data of a particle track and computes a B-spline curve for the track. This step includes noise reduction and allows computing 1st and 2nd order derivatives for velocity and acceleration at any position in time within the time interval of the observed particle track.
\item \emph{flowfit} takes particle locations and any other physical quantity for each such location such as velocity or acceleration for one particular point in time and computes a 3D B-spline curve that optionally satisfies other constraints such as freedom of divergence or curl. This step involves solving a linear weighted least squares problem.
\end{itemize}

These steps share many similarities. Both make use of B-splines to represent the result as a continuous function and both employ a similar form of noise reduction via penalization like it was introduced in \cite{eilers96}. The difference is that \emph{trackfit} deals with data that is already equidistantly sampled and one-dimensional while \emph{flowfit} deals with scattered data in three dimensions. \emph{flowfit} is also able to compute vector fields that are free of divergence or curl by extending the equation system that is used for computing the B-spline weighting coefficients with appropriate equations penalizing divergence or curl. This is useful for incompressible flows and improves the spatial resolution to some extent.

\section{B-splines}

Instead of restricting ourselves to time and space discrete signals for a particle track or a velocity field we can try to reconstruct a \emph{continuous} function with a finite number of degrees of freedom. Building a function such as a particle track that maps time to particle location or velocity field that maps location to flow speed as a weighted sum of B-splines is one option to model a continuous function and has several benefits: The function will automatically be as smooth as one desires and function evaluation including temporal or spatial derivatives is fast. For example, a cubic B-spline is twice continuously differentiable and so is every linear combination of cubic B-splines. Also, such a representation allows us to express the function's value or any derivative at any point exactly as a linear combination of weights without the need to numerically approximate derivatives.

Throughout this paper the $k$-th order cardinal B-spline function centered at zero will be referred to as $\beta_k$. This family of functions can be defined recursively in the following way:

\begin{equation} \label{eq:bsplinek}
\begin{array}{rcl}
\beta_k & : & \mathbb{R} \longmapsto \mathbb{R} \\
\beta_0(x) & = & \begin{cases}
                   1           & \text{for } |x| < \frac{1}{2} \\
                   \frac{1}{2} & \text{for } |x| = \frac{1}{2} \\
                   0           & \text{else}
                 \end{cases} \\
\beta_k(x) & = & \frac{1+k+2x}{2k} \beta_{k-1}(x + \frac{1}{2}) + \\
           &   & \frac{1+k-2x}{2k} \beta_{k-1}(x - \frac{1}{2})
\end{array}
\end{equation}

\noindent In the special case $k=2$ for a quadratic B-spline, this can be written as

\begin{equation} \label{eq:beta2}
\beta_2(x) = \begin{cases}
        \frac{3}{4} - |x|^2 & \text{for } |x| < \frac{1}{2} \\
        \frac{1}{2} \left( \frac{3}{2} - |x| \right)^2 & \text{for } \frac{1}{2} \leq |x| < \frac{3}{2} \\
        0 & \text{for } \frac{3}{2} \leq |x|
             \end{cases}
\end{equation}

\noindent and for the special case $k=3$, a cubic B-spline, we get

\begin{equation} \label{eq:beta3}
\beta_3(x) = \begin{cases}
               \frac{4}{6} - |x|^2 + \frac{1}{2} |x|^3 & \text{for } |x| < 1 \\
               \frac{1}{6} \left( 2 - |x| \right)^3    & \text{for } 1 \leq |x| < 2 \\
               0                                       & \text{for } 2 \leq |x|
             \end{cases}
\end{equation}

\section{Filtering particle tracks}

A time-discrete 3D particle track can be viewed as three digital signals that represent how the particle's x-, y- and z coordinates change over time. A simple but reasonable model of the measured particle locations is that they are the sum of the particle's real locations and a measurement error signal. Typically, the high frequency portion of these measured location signals are dominated by measurement noise while in the low frequency portion the measurement noise will be negligible compared to the signal. One way of dealing with this is to use a corresponding Wiener filter that has the goal of minimizing the sum of squared errors.

\begin{figure*}
  \centering
    \includegraphics[scale=0.85]{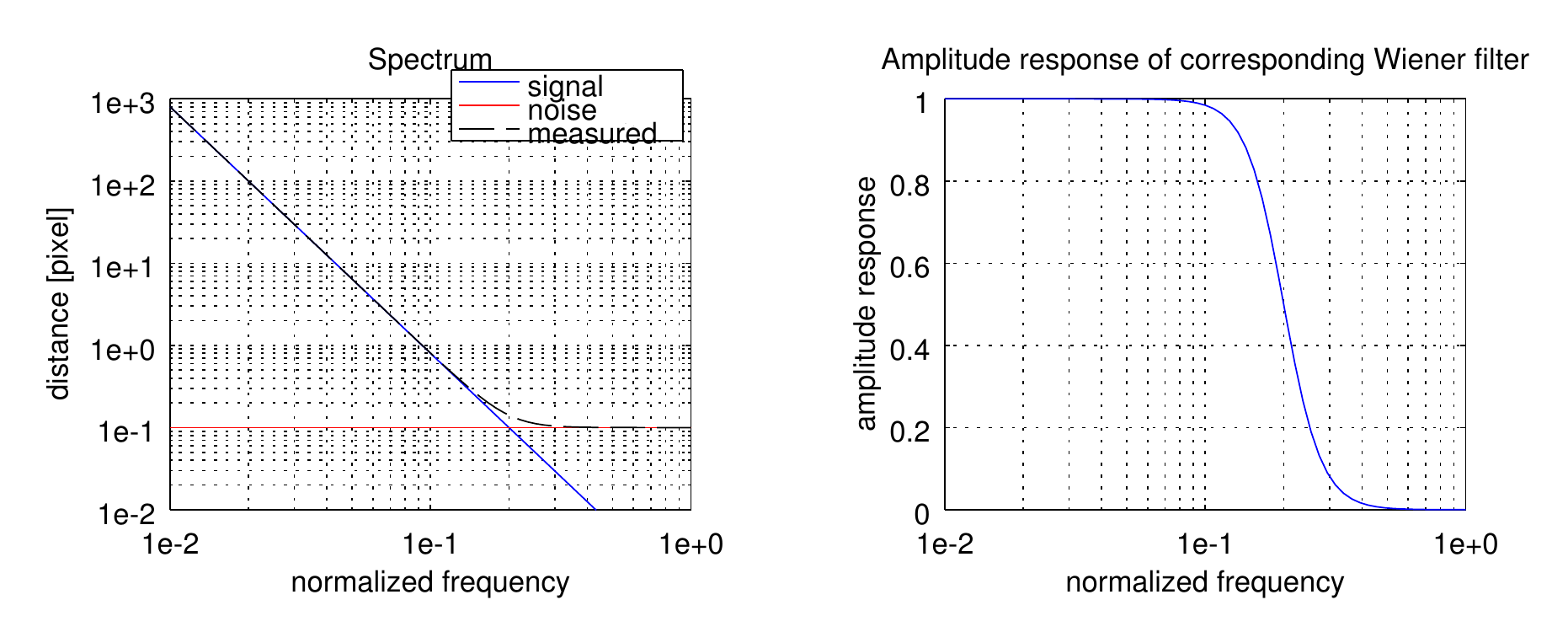}
  \caption{Signal and noise model along with the corresponding Wiener filter}
  \label{fig:wienerspectrum}
\end{figure*}

Under the assumption that the measurement noise is not correlated with the true particle's locations the optimal Wiener filter simplifies to a filter with a phase response of constant zero and an amplitude response that is completely determined by the signal-to-noise ratio in the following way where $f$ refers to the frequency and $SNR$ refers to the power spectral density ratio between the noise-free signal and the noise:

\begin{equation} \label{eq:snr2ampresponse}
A(f) = \frac{SNR(f)}{SNR(f) + 1}
\end{equation}

After performing a spectral analysis of particle track data of different flow experiments we observed the following: The square root of the true particle locations' power spectral density had a shape roughly proportional to $f^{-3}$ near the the frequency where the signal-to-noise ratio equals one. The noise floor was flat which is expected when location measurement errors of different points in time don't correlate with each other. This leads to our simple model of the signal and noise spectra for which a specific Wiener filter follows using equation \ref{eq:snr2ampresponse}, see figure \ref{fig:wienerspectrum}.

After experimenting with different kinds of filter design methods and implementations we noticed that such a Wiener filter can be approximated well using a B-spline fit with the appropriate choice of penalization. The input to \emph{trackfit} are $n$ measured values $y_i$ for $1 \leq i \leq n$, representing for example all the y-components of a particle that was tracked for $n$ equidistantly spaced points in time. Without loss of generality we can assume a time step of one. With knots at locations $0, 1, \ldots, n+1$, a cubic spline function for the interval $[1, n]$ can be represented as the following weighted sum of $n+2$ B-splines

\begin{equation} \label{eq:splinefunc}
\vec{p_c}(t) = \sum_{i=0}^{n+1} \vec{c_i} \beta_3(t - i)
\end{equation}

\noindent where $c_i$ are the unknown weighting coefficients and $\beta_3$ refers to the cubic cardinal B-spline defined in the previous section.

We compute these coefficients by minimizing the cost function

\begin{equation} \label{eq:trackfitcost}
\begin{array}{rl}
F(c) = & \sum_{i=1}^n \left| p_c(i) - y_i \right|^2 + \\
       & \sum_{i=1}^{n-1} \left| \lambda \left( c_{i-1} - 3 c_i + 3 c_{i+1} - c_{i+2} \right) \right|^2
\end{array}
\end{equation}

\noindent where the parameter $\lambda$ controls how strongly the third order finite differences are penalized in relation to the error between measurement and the fitted curve. A very large value for $\lambda$ approaching infinity will result in a spline curve that approaches a quadratic polynomial for the whole particle track. Smaller values will lead to a spline curve that will follow the measurement data more closely. We can also interpret this approach as a Kalman-like filter: The measured particle locations are combined with a physical model in which the change in acceleration is assumed to be white noise and the parameter $\lambda$ is chosen as ratio between the standard deviation of the location measuement noise and the standard deviation of the unpredictable change in acceleration in order to estimate the most likely particle track under this model.

The resulting mathematical problem is weighted linear least squares problem with a sparse matrix. The matrix of the corresponding normal equation system, see equation \ref{eq:trackfitnormaleqs}, will be a symmetric positive definite 7-band matrix for which efficient factorizations can be computed in-place to solve for the B-spline weights $c_i$ directly. The condition of the normal equation system is acceptible for typical machine accuracies and choices of $\lambda$.

\begin{equation} \label{eq:trackfitnormaleqs}
A c = b
\end{equation}

We refer to the frequency at which the power spectral density of the signal and noise cross each other the cutoff frequency. The optimal Wiener filter would have an amplitude response of $\frac{1}{2}$ at this frequency which follows from equation \ref{eq:snr2ampresponse}. For a normalized cutoff frequency $f_{cutoff}$ between $0.1$ and $0.5$ where $1$ represents the Nyquist frequency choosing $\lambda$ according to equation \ref{eq:trackfitlambda}

\begin{figure}
  \centering
    \includegraphics[width=\columnwidth]{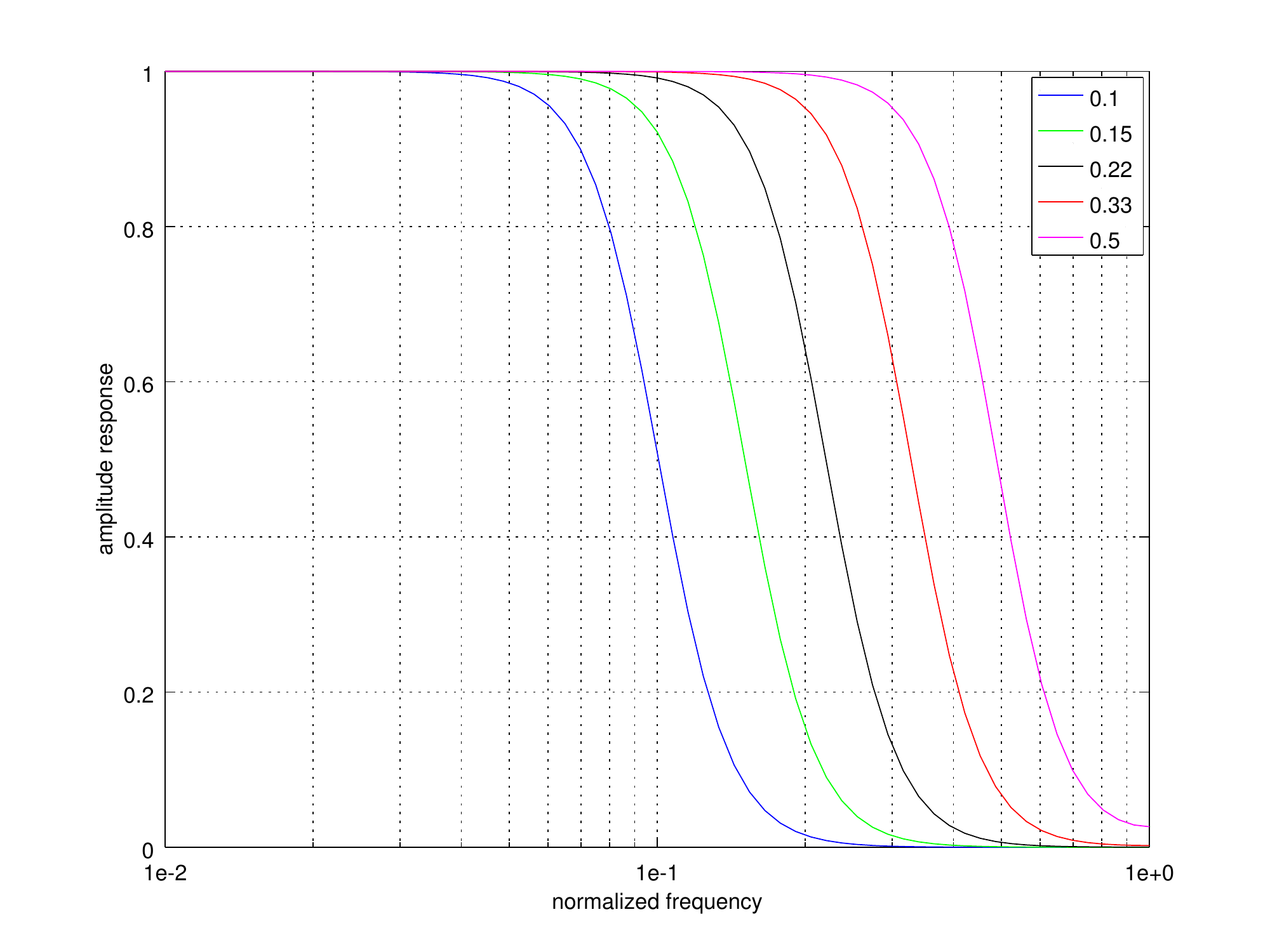}
  \caption{Amplitude responses of different example filters with varying cutoff frequencies}
  \label{fig:trackfit_filters}
\end{figure}

\begin{equation} \label{eq:trackfitlambda}
\lambda = \left(\frac{1}{\pi \cdot f_{cutoff}}\right)^3
\end{equation}

\noindent will result in a a B-spline fit that approximates the optimal Wiener filter quite well, compare figures \ref{fig:wienerspectrum} and \ref{fig:trackfit_filters}. The cutoff frequency is in the denominator so that a lower frequency will lead to a stronger penalization of the third order derivative. The third power is due to the fact that we are penalizing the third oder derivative which scales with the third power of a frequency.

To determine the magnitude response of the filter we compute the Fourier transform of an impulse response. To compute the impulse response we set $n$ to 201, $y_i$ to zero for all $i \neq 101$ and $y_{101}$ to one. Figure \ref{fig:trackfit_filters} shows five examples with different values for $\lambda$ for different cutoff frequencies.

After computing the B-spline weighting coefficients $c$ for a specific particle track, the track curve can be evaluated at any point in time along with its first and second derivatives. This allows temporal super sampling of the curve with consistent particle locations, velocities and accelerations for later processing.

It is worth pointing out that under the assumption our particle track model holds and the $\lambda$ parameter was chosen correctly, variances of the B-Spline coefficient errors due to the measurement noise will be larger at the borders of the track than at center of the track. This can be verified by inspecting the diagonal of the inverse of $A$ from equation \ref{eq:trackfitnormaleqs} which represents all the variances of $c$ multiplied by $2n-1$ times the variance of the particle location measurement noise. This means that one should not put equal trust in the computed locations, velocities and accelerations over the complete time interval.

\section{Reconstructing a 3D vector field}

For the spatial reconstruction of a vector field given scattered data for a particular point in time, it is possible to extend the method of penalized B-splines to multiple dimensions on a Cartesian lattice with a point distance $h$. For every lattice point within a certain cube we would have $d$ degrees of freedom, for example, $d=3$ for a velocity or material acceleration field, that are the weighting factors for the corresponding 3D B-spline functions of order $k$.

Suppose $l \in \mathbb{R}^3$ is the coordinate of the lower corner lattice point of the volume and $N \in \mathbb{N}^3$ describes the number of grid points in each dimension. Then, we can assign each grid point an index between $1$ and $n = N_1 N_2 N_3$ along with a world coordinate $x_i \in \mathbb{R}^3$ for the $i$-th lattice point derived from $l$ and $h$ and represent the vector field as the function shown in \ref{eq:flowfit_curve_model}

\begin{equation} \label{eq:flowfit_curve_model}
\begin{array}{rcl}
\vec{v} &:& \prod_{j=1}^3[l_j + \frac{k-1}{2} h, l_j + \left( N_j - \frac{k+1}{2} \right) h] \longmapsto \mathbb{R}^3 \\
\vec{v}(\vec{x}) &=& \sum_{i=1}^{n} 
  B_k\left( \frac{x - x_i}{h} \right) \vec{c_i}
\end{array}
\end{equation}

\noindent Here, $B_k$ is a 3D convolution of separate one-dimensional B-splines $\beta_k$:

\begin{equation} \label{eq:bigbeta}
\begin{array}{rcl}
B_k    & : & \mathbb{R}^3 \longmapsto \mathbb{R} \\
B_k(x) & = & \prod_{i=1}^3 \beta_k(x_i)
\end{array}
\end{equation}

\noindent With this model of the vector field each given data point results in $d$ linear equations involving $d \left( k + 1 \right)^3$ unknown B-spline weighting coefficients because for any such point there are $k+1$ lattice points for each spatial dimension that contribute to the value of the resulting function and we have $d$ separate variables for each lattice point.

In addition to equations for the data points it is possible to add regularizations for overall smoothness (similar to how a higher order derivative is penalized in the \emph{trackfit} approach) but also to penalize other physical properties such as the divergence of a velocity field or the rotation of a material acceleration field assuming that acceleration is dominated by the pressure gradient. Both the divergence and rotation of the vector field at an arbitrary point in the domain of $\vec{v}$ can be expressed as a linear combination of the unknown variables $\vec{c_i}$ so that resulting optimization problem is still a linear least squares problem.

Choosing the B-spline order $k$ is a trade-off between how smooth the function and how sparse the matrix of the resulting equation system should be. 

This paper is incomplete. Please wait for an updated version.


\begin{thebibliography}{9}

\bibitem{eilers96}
  Eilers, P.H.C. and Marx, B.D. (1996).
  {\em Flexible smoothing with B-splines and penalties}.
  Statistical Science 11(2): 89-121.

\end{thebibliography}
\end{document}